\documentstyle[aps,twocolumn]{revtex}
\begin{document}

\noindent{\bf Fendley, Ludwig and Saleur reply to Skorik's comment:}

In a series of papers \cite{FLS}, we have used a novel approach,
combining the Bethe ansatz with a kinetic (Boltzmann) equation, in
order to compute exactly transport properties in a Luttinger liquid
with an impurity (also known as the boundary sine-Gordon model [BSGM]).
In a recent comment \cite{yuk}, Skorik claimed that while our results
were correct in linear response, there was a ``serious flaw'' in our
calculation at finite voltage. We explain here why Skorik's arguments
are inappropriate, and that there is no flaw in our work.

In our approach, the effect of the voltage corresponds to injecting
into the sample left movers and right movers from different
reservoirs, at chemical potentials, $\mu_{L,R}=\pm {eV\over 2}$
respectively\cite{KF,ACF}.  We use a Landauer-B\"uttiker scattering
approach \cite{LB}.  By a non-linear change of basis called
``folding'' \cite{FLS}, we map this to a problem with reservoirs
setting different chemical potentials for the quasiparticles of {\bf
charge $\pm $}. These quasiparticles are\cite{FSW} the ``massless
limit'' of the usual sine-Gordon soliton and antisoliton states.
Because the model is integrable, these scatter one-by-one off each
other and off the impurity, conserving their momentum and either
conserving or switching charge, with exactly known reflection and
transmission $S$-matrices.  This permits the construction of the exact
(scattering) eigenstates of the {\it interacting} Hamiltonian,
describing the Luttinger liquid leads {\it plus} the impurity, in a
way analogous to ordinary potential scattering.  To compute the
conductance, we use a Boltzmann equation, counting how much charge is
transported through the impurity in the presence of different
populations of $\pm$ quasiparticles, set by the bias. This is natural,
in spite of the interacting nature of the problem, because of the very
simple nature of collisions that follows from integrability.  In fact,
we note that subsequent to our original work, the Boltzmann equation,
in the case of linear response, was derived\cite{LSII} directly from
the Kubo formula combined with form-factors, i.e.\ matrix elements of
current operators in the quasi particle basis, and complete agreement
with the earlier results\cite{FLS} was found.

The use of the Boltzmann equation {\bf per se} was not criticized by
Skorik.  As far as we understand it, his concern is that in the
Boltzmann equation we used scattering matrix elements ``at zero
voltage'', whereas, for finite voltage, the $S$ matrix itself might
potentially acquire a voltage dependence, determined by the $V$
dependent filling of the ground state. We will show that this is in
fact {\it not} the case.  First, we note that following Skorik's
logic, the same criticism could seemingly be made for the
linear-response calculation at non-zero temperature. Here, the ground
state is not shifted by the voltage, but physical properties do not
depend so much on the zero-temperature ground state as on the states
whose filling fractions are the ones of thermal equilibrium (see
\cite{LLSS} for more details on this). Nevertheless, Skorik agrees
that the zero-temperature $S$ matrix is the appropriate one to use
here, and that it does not acquire any sort of temperature dependence.
As discussed in detail previously \cite{FLS}, the only effect of the
temperature on the Boltzmann equation is the appearance of the
non-trivial filling fractions. To use another language, we are still
describing the problem in terms of the ``bare'' particles (we mean
bare in the sense of thermal fluctuations, not quantum fluctuations).
The integrability means that these bare particles are a valid basis
for the problem and that they still scatter one-by-one off each other
and off the impurity despite the macroscopic number of particles
excited around a given particle at non-zero temperature. (More
precisely, this means that we do calculations in the extremely dilute
limit where a particle description is valid; the central assumption of
this and all thermodynamic Bethe ansatz computations is that no phase
transition interferes with the continuation of the result to the
regime of finite densities.)

Consider now a non-zero voltage, and let us discuss the simplest case
of $T=0$.
The Fermi sea  is, indeed, filled with {\it bare} (with respect
to the zero-voltage sea) quasiparticles, as discussed by Skorik.
However,  each  of these
 scatters with
 the same, field-independent $S$-matrix on the impurity. It is this
$S$ matrix
that is used to build the asymptotic states, it is therefore the  one
that
appears in the Boltzmann equation  in \cite{FLS}. Of course, one can
be interested in  the $S$-matrix
for scattering particles excited {\it on top} of the finite-field
sea:
by scattering these particles through the impurity and the sea, one
gets then
a field dependent result. A  computation of that type was performed
for the  the Kondo problem in a magnetic field in \cite{A}, as
mentioned by Skorik.
Such a dressed $S$ matrix  has actually appeared in other
computations we made, such as low frequency AC properties \cite{LSI}.
 But as far as the DC conductance is concerned, the only object
necessary is the one to build asymptotic states, and it is the ``bare
$S$ matrix'' as used in \cite{FLS}. In fact, even if one wanted to
use the ``dressed'' $S$ matrix in our Boltzmann equation, nothing
would change. This is because, at the reflection less points of the
sine-Gordon model, to which we have restricted, the bulk scattering
is diagonal, so the dressing is a mere {\it phase}  (it is also   the
 phase that is discussed in
\cite{A}.)  Since in our Boltzmann-equation approach the DC current
depends only on probabilities, i.e.\ modulus squared of $S$ matrix
elements,  the $V$-dependence of this $S$-matrix would  drop out
anyway. This, we think, invalidates Skorik's criticism.

There is one major difference between non-zero temperature and
non-zero voltage, which might have created some confusion:
  As emphasized correctly  by Skorik, diagonalizing
the BSGM
directly at nonzero voltage
is a difficult exercise,
 since the boundary interaction does
not conserve the charge. It was not done in \cite{FLS}: there, we
argued
instead  that, physically,   the role of the voltage is
 to fix the populations
of quasiparticles; it is applied far from the impurity, and {\bf
hence},
since the excitations are localized solitons, in a region where the
effect
of the impurity is negligible. Thus at least when computing DC
transport properties through the impurity, the effect of the voltage
is  to shift the Fermi sea but does not interfere with the one-by-one
scattering off the impurity (this is actually one of the basic ideas
underlying
the Landauer-B\"uttiker approach).

In fact, it is also possible to diagonalize the BSGM with a voltage,
and to confirm this physical argument \cite{FLeSII,LSI,LSII,BLZ}. The
main idea  is
to observe that the physical properties
of the BSGM
  are the same as the properties of
another model, where the boundary interaction $\cos\phi$ is replaced
by another interaction of the Kondo type
$S^+e^{-i\phi}+S^-e^{i\phi}$,
where the spin is taken in,
either a cyclic representation of the
$U_qsl(2)$ algebra \cite{FS}, or a representation of the oscillator
algebra \cite{BLZ}. The key property of these representations is that
all monomials (of total vanishing charge) in $S^\pm$ have the same
trace.  One can then compare the perturbative computations of
properties like the partition function, or the conductance using the
Keldysh formalism \cite{CFWI} in the two models, and prove their
equivalence (this method is used in \cite{CFWII} in the case
$g={1\over 2}$, where the $S^\pm$ are equivalent to boundary
fermions).  The advantage of this new formulation is that there is a
conserved charge, the sum of the quasiparticles charge and the spin
$S^z$ of the boundary degree of freedom.  The voltage is then
included
by shifting the field $\phi\to \phi+gVt$ in the boundary interaction
\cite{KF,CFWI,CFWII}. As discussed in the appendix of \cite{LSII},
this is equivalent to not shifting $\phi$, but applying a magnetic
field on the boundary spin, $h=gV$.  Still in the appendix of
\cite{LSII}, it is explained how the problem with a field applied
only
to the boundary spin, and the problem with a field applied to the
  spin {\bf and} another field applied to the bulk $U(1)$
charge are related by a unitary transformation (this is closely
related to the behavior of the electrons-impurity susceptibility in
the Kondo problem, as discussed recently in \cite{AB}. This
transformation shifts the overall charge by a constant, but does not
modify difference of $U(1)$ charges). We can then consider the
problem
where the field is coupled to the conserved charge, which is of
course
trivial to diagonalize.  Asymptotic states can then be written
explicitly for this auxiliary problem with impurity {\bf and}
voltage; they involve of course the same S matrix as the ones without
voltage (the same occurs in the $g={1\over 2}$ case\cite{CFWII}).
Scattering eigenstates in the traditional sense can be constructed,
and the Landauer-B\"uttiker approach can then be applied
to compute the current, the DC fluctuations \cite{FSII}, and some of
the AC properties \cite{LSI}.

In conclusion, the formula proposed in \cite{FLS} does
not
suffer from Skorik's criticisms. For the reader who does not want to
follow the detailed arguments presented above, we observe that our
formula has passed successfully more
tests than recognized in Skorik's comment. In addition to the case
free-fermion case $g={1\over 2}$, our
formula reproduces the  correct
 result for $g\to 0$ and $g=1$. Even though the
latter case consists of free fermions in the unfolded version
of the problem, {\bf it } is highly non trivial in our folded point
of view:
indeed, the solitons and antisolitons scatter then with an $SU(2)$
invariant S matrix, identical to the one in the Kondo problem. The
former case,
while simple to study
in the classical limit, is highly non-trivial  from the point of view
of
integrability: it involves
taking the complicated limit of an interacting quantum problem, where
an infinity of quasiparticles scatter, all
with non trivial S matrices, and the filling fractions of the
solitons and antisolitons
are far from simple. The final result for the $g\to 0$ limit,
expressed in
terms of Bessel functions of imaginary
arguments) \cite{BLZ,FFS}, agrees with the result obtained using
a Fokker-Planck equation \cite{CF}.
In addition, for any $g$ our formula displays the right behavior in
the
strong and weak backscattering limits,
the existence
of a maximum for large enough voltage which is expected on physical
grounds \cite{FLS}. Finally, it also obeys a duality (proven at $T=0$
in
\cite{FLS} and related
with very natural analyticity conjectures at $T>0$ in \cite{BLZ})
between weak and strong backscattering, where
the appropriate tunneling particles are Laughlin quasiparticles and
electrons
respectively. This duality, while not established rigorously  prior
to our work,
is considered highly desirable
on physical grounds \cite{CFWI,CFWII}. We thus see no reason to cast
doubts on our result, and hope that sooner or later, it will be
compared favorably with numerical simulation or experimental  data.

\bigskip

\noindent P. Fendley$^1$, A.W.W. Ludwig$^{2}$ and H. Saleur$^{3}$
\medskip

\noindent$^1$ Department of Physics, University of Virginia,
Charlottesville, VA 22901

\noindent$^2$ Physics Department, University of California,
Santa Barbara, CA 93106

\noindent$^3$ Department of Physics, University of Southern
California,
Los Angeles CA 90089-0484

\end{document}